\documentclass[useAMS,usenatbib]{mn2e}

\usepackage{amssymb,alltt}
\usepackage{natbib}
\usepackage{graphicx}
\usepackage{xspace}
\usepackage{url}
\usepackage{hyperref}
\usepackage{wasysym}
\usepackage[T1]{fontenc}
\usepackage{aecompl}

\def\u{{\bf u}}

\def\thetav{{\bf \theta}}

\title{Weak Lensing Galaxy Cluster Field Reconstruction}
\author[E.~Jullo,  S.~Pires, M.~Jauzac \& J.-P. Kneib]
{E.~Jullo$^1$,
  S.~Pires$^2$, 
 M.~Jauzac$^3$ \& 
  J.-P.~Kneib$^{4,1}$ \\
  $^1$Aix Marseille Universit\'e, CNRS, LAM (Laboratoire d'Astrophysique de Marseille) UMR 7326, 13388, Marseille, France  \\
  $^2$Laboratoire AIM, CEA/DSM-CNRS, Universit\'e Paris 7 Diderot, IRFU/SAp-SEDI, Service d'Astrophysique, CEA Saclay, Orme des \\Merisiers, 91191 Gif-sur-Yvette, France\\
  $^3$Astrophysics and Cosmology Research Unit, School of Mathematical Sciences, University of KwaZulu-Natal, Durban 4041, South Africa\\
  $^4$LASTRO, Ecole polytechnique f\'ed\'erale de Lausanne, Suisse}
\date{Released 2013 Xxxxx XX}

\pagerange{\pageref{firstpage}--\pageref{lastpage}} \pubyear{2012}

\begin{document}

\label{firstpage}

\maketitle

\begin{abstract}
In this paper, we compare three methods to reconstruct galaxy cluster density fields with weak lensing data. The first method called FLens integrates an inpainting concept to invert the shear field with possible gaps, and a multi-scale entropy denoising procedure to remove the noise contained in the final reconstruction, that arises mostly from the random intrinsic shape of the galaxies. The second and third methods are based on a model of the density field made of a multi-scale grid of radial basis functions. In one case, the model parameters are computed with a linear inversion involving a singular value decomposition. In the other case, the model parameters are estimated using a Bayesian MCMC optimization implemented in the lensing software Lenstool. Methods are compared on simulated data with varying galaxy density fields. We pay particular attention to the errors estimated with resampling. 
We find  the multi-scale grid model optimized with MCMC to provide the best results, but at high computational cost, especially when considering resampling. The SVD method is much faster but yields noisy maps, although this can be mitigated with resampling. The FLens method is a good compromise with fast computation, high signal to noise reconstruction, but lower resolution maps. 
All three methods are applied to the MACS J0717+3745 galaxy cluster field, and reveal the filamentary structure discovered in \citet{ejlens:jauzac12}. We conclude that sensitive priors can help to get high signal to noise, and unbiased reconstructions.
\end{abstract}

\section{Introduction}

Galaxy redshift surveys such as SDSS \citet{ejlens:york00} and N-body simulations of cosmic structure formation (for example the Millenium simulations \citet{ejlens:springel05}) have revealed a complicated network of matter, in which massive galaxy clusters are located at the nodes, filaments connect them to each others, and in-between extended regions with few galaxies and matter called voids fill about 80\% of the volume of the Universe \citep{ejlens:pan12,ejlens:bos12}.

Galaxy clusters are of considerable cosmological interest, as they are the most recent structures to have formed at the largest angular scales. Taking advantage of this specificity, several cluster-related cosmological probes have been developed either based on cluster count statistics \citep{ejlens:berge08,ejlens:pires09,ejlens:shan12} or on the study of their physical 
properties (e.g. triaxialityÊ\citet{ejlens:morandi11}, bulleticity \citet{ejlens:massey11} or  gas mass fraction \citet{ejlens:rapetti10}).

Filamentary structures surrounding galaxy clusters also happen to be of particular interest. On the one hand, they reveal cosmological voids and alike cluster count statistics, void number counts and sizes are effective cosmological probes  (\citet{ejlens:davis11,ejlens:higuchi12,ejlens:krause13}). On the other hand, filaments funnel matter onto the galaxy clusters, and as such they play an important role in cluster and galaxy formation.

Lensing has recently demonstrated its effectiveness at mapping filaments. For instances, \citet{ejlens:heymans08} has uncovered a filamentary structure between the pair of clusters 901 and 902. In their analysis of the double cluster system Abell 222 and Abell 223, \citet{ejlens:dietrich12} showed evidence for a possible dark matter filament connecting both clusters. Finally, in the COSMOS field \citep{ejlens:scoville07}, \citet{ejlens:massey07} uncovered a massive large-scale structure at redshift $z\sim0.73$ extending over about 1 degree in length. 

Recently, \citet{ejlens:jauzac12} claimed another detection of a large-scale filament connected on one end to the massive cluster MACS J0717+3745, and vanishing into the cosmic web on the other end. They used a model made of a multi-scale grid of radial basis functions (RBF) and a Bayesian MCMC optimization algorithm implemented in the lensing software Lenstool to map its mass distribution and measure its size and density. 

In this paper, we study three methods of lensing map reconstruction, including the method used in \citet{ejlens:jauzac12}. The first method called FLens integrates an inpainting concept to invert the shear field with possible gaps, and a multi-scale entropy denoising procedure to remove the noise contained in the galaxies. The second and third method are based on the same model of multi-scale grid of RBFs, but in one case the parameters are estimated with Lenstool, and in the other case with a linear matrix inversion involving a singular value decomposition. We use simulated data and compare the reconstructed maps in terms of fidelity to the input map, sensitivity to the density of galaxies in the input weak lensing catalog. We also pay particular attention to the errors estimated either directly from the MCMC samples or the linear inversion theory, and the errors estimated with resampling. 

The outline of the paper is the following. In \S \ref{sec:method}, we review the formalism of the different techniques. In \S \ref{sec:simus}, we use simulations to compare the methods, focusing successively on the reconstructing maps, azimuthally averaged density profiles, errors and signal to noise maps. Finally in \S \ref{sec:macs0717}, we compare the reconstructions obtained with the different methods applied to real data coming from HST observations of the massive galaxy cluster MACS J0717+3745. Throughout this paper, we compute cosmological distances to lensed galaxies assuming the Universe is flat and described by the $\Lambda$CDM model with $\Omega_mÊ= 0.3$ and $w = -1$.

\section{Methods}
\label{sec:method}

\subsection{Weak Lensing formalism}

Gravitational lensing i.e. the process by which light from distant galaxies is bent by the gravity of intervening mass in the Universe, is an ideal tool for mapping the mass distribution of lensed structures because it depends on the total matter distribution of the intervening structures.

In lensing, the spin-2 shear field ${\bf \gamma_i}(\thetav)$ that is derived from the shapes of observed background galaxies, can be written in terms of the intervening lensing gravitational potential $\psi(\thetav)$ projected on the sky  \citep{ejlens:bartelmann01}:
\begin{equation}
\label{gamma}
\begin{array}{l} \gamma_1(\theta)= \frac{1}{2}(\partial_1^2-\partial_2^2)\psi(\thetav)\\ \gamma_2(\thetav)=\partial_1\partial_2\psi(\thetav), \end{array}
\end{equation}
where the partial derivatives $\partial_i$ are with respect to $\theta_i$. 

The convergence $\kappa(\thetav)$ can also be expressed in terms of the lensing potential $\psi(\thetav)$,
\begin{equation}
\label{kappa}
\kappa(\thetav)=\frac{1}{2}(\partial_1^2 + \partial_2^2) \psi(\thetav),
\end{equation}
and is related to the mass density $\Sigma(\theta)$ projected along the line of sight by
\begin{equation}
\label{sigma1}
\kappa(\thetav) = \frac{\Sigma(\thetav)}{\Sigma_{crit}},
\end{equation}
where the critical mass density $\Sigma_{crit}$ is given by
\begin{equation}
\label{sigma2}
\Sigma_{crit}=\frac{c^2}{4\pi G}\frac{D_{OS}}{D_{OL}D_{LS}},
\end{equation} 
where $G$ is Newton's constant, $c$ the speed of light, and $D_{OS}$, $D_{OL}$, and $D_{LS}$ are the angular-diameter distances between the observer (O),  the lens (L),  and a galaxy source  (S) at an arbitrary redshift.

\subsection{A new inverse method}

If the shear field could be measured everywhere, the convergence field could be determined without error. In reality, we only have  access to an estimator of the shear field at the random discrete locations of the background galaxies.
The shear information is contained in the observed ellipticity of the background galaxies, but is  overwhelmed by the intrinsic galaxy own ellipticity. Fortunately, we can assume that this intrinsic shape noise is random and Gaussian distributed. Therefore we can compute an unbiased estimate of the shear by binning the galaxies in a grid and average their ellipticities.

\subsubsection{The Kaiser \& Squires inversion}
The weak lensing mass inversion problem consists in reconstructing the projected (normalized) mass distribution $\kappa(\theta)$ from the measured shear field $\gamma_i(\theta)$ in a grid. We invert Eq. (\ref{gamma}) to find the lensing potential $\psi$ and then apply formula Eq. (\ref{kappa}) to obtained  $\kappa(\theta)$. This classical method is based on the pioneering work of \citet[][KS93]{ejlens:kaiser93}. In short, this corresponds to :
\begin{eqnarray}
\label{eqn_reckE}
\tilde \kappa  & = & \Delta^{-1}\left((\partial_1^2 - \partial_2^2) \gamma_1+ 2 \partial_1\partial_2 \gamma_2\right)  \nonumber  \\
   & =  & \frac{\partial_1^2 - \partial_2^2}{\partial_1^2 + \partial_2^2} \gamma_1+ \frac{2 \partial_1\partial_2}{\partial_1^2 + \partial_2^2}\gamma_2.
\end{eqnarray}

Taking the Fourier transform of these equations, we obtain 
\begin{equation}
\hat{\kappa} = \hat P_1 \hat{\gamma_1} + \hat P_2 \hat{\gamma_2},
\end{equation}
where the hat symbol denotes Fourier transforms and we have defined
$k^2 \equiv k_1^2 + k_2^2$ and
\begin{eqnarray}
\hat{P_1}(\mathbf k) & = & \frac{k_1^2 - k_2^2}{k^2} \nonumber \\
\hat{P_2}(\mathbf k) & = & \frac{2 k_1 k_2}{k^2},
\end{eqnarray}
with $\hat{P_1}(k_1,k_2) \equiv 0$ when $k_1^2 = k_2^2$, and
$\hat{P_2}(k_1,k_2) \equiv 0$ when $k_1 = 0$ or $k_2 = 0$.

Note that to recover $\kappa$ from both $\gamma_1$ and $\gamma_2$, there is a degeneracy when $k_1 = k_2 = 0$. 
Therefore, the mean value of $\kappa$ cannot be recovered from the shear maps. This is known as the mass-sheet degeneracy.
This problem can be solved with additional information such as lensing magnification measurements for instance.

In reality, the measured shear is noisy because only a finite number of galaxy ellipticities are  averaged per pixel. The actual relation between the measured shear $\gamma_{ib}$ in pixel $b$ of area $A$ and the true convergence $\kappa$   is 
\begin{equation}
\label{eq_gamma}
\gamma_{ib} = P_i * \kappa + n_i\;,
\end{equation}
where the intrinsic galaxy shape noise contribution $n_i$ is Gaussian distributed with zero mean and width $\sigma_n \simeq \sigma_\epsilon/\sqrt{N_g}$. The average number of galaxies in a pixel $N_g = n_g\ A$ depends on the the average number of galaxies per square arcminute  $n_g$. 
The ellipticity dispersion per galaxy $\sigma_\epsilon$ arises both from measurement errors and the dispersion in the intrinsic shape of galaxies. 

From the central limit theorem, we can assume to a good approximation that with $n_g \simeq 10$ galaxies per square arcminute, in pixels with area $A \gtrsim 1$ square arcminute the noise $n_i$ is Gaussian distributed and uncorrelated.



The most important drawback of the KS93 method is that it 
requires a convolution of shears to be performed over the entire sky. As a result, if the field is small or irregularly-shaped, then the method can produce artifacts 
in the reconstructed matter distribution near the boundaries. 

\subsubsection{The Seitz \& Schneider inversion}
In \citet{ejlens:seitz96}, the authors propose a local inversion method that reduces these unwanted boundary effects. The convergence $\kappa$ is computed in real space (without Fourier transform) thanks to the kernel integration

\begin{equation}
\kappa(\theta)-\kappa_0=\frac{1}{\pi}\int_{\theta' \in \Omega}K(\theta - \theta')\cdot {\bf \gamma}(\theta')\, d\theta',
\end{equation}
where $\kappa_0$ stands for the mean value of $\kappa$. The kernel $K$ depends on the geometry of the domain $\Omega$. For $\Omega=\mathbb{R}^2$, it is given by

\begin{equation}
K(\theta)=\left(\frac{\theta_2^2-\theta_1^2}{(\theta_1^2+\theta_2^2)^2} , \frac{-2\theta_1\theta_2}{(\theta_1^2+\theta_2^2)^2}\right) .
\end{equation} 

\noindent where we expressed the positions in complex coordinates $\theta = \theta_1 + i\theta_2$. For small irregularly-shaped fields, the authors propose to combine the derivatives of ${\gamma_i}$
\begin{equation}
\u=\left(\begin{array}{l} \partial_1\gamma_1+\partial_2\gamma_2 \\ \partial_1\gamma_2-\partial_2\gamma_1 \end{array}\right),
\end{equation}
and then to apply the Helmholtz decomposition $\u=\nabla\kappa^{(E)}+\nabla\times\kappa^{(B)}$, in order to reconstruct the convergence $\kappa=\kappa^{(E)}$. 
This method reduces the unwanted boundary effects but whatever 
the formula, the reconstructed field is more noisy than that one obtained 
with a global inversion. Another point is that the reconstructed dark matter 
mass map  still has a complex geometry that will complicate subsequent analyses.

\subsubsection{The FLens method}
\label{sec:flens}

{\bf Binning the shape catalogue}\\
As said previously, the shape catalogue is first binned into a regular grid, in which  each pixel value is obtained by averaging the ellipticity of the galaxies it contains. The pixel size is a parameter defined by hand, so that all (or almost all) pixels contain at least one galaxy. Not doing so usually prevents mass inversion because of missing data. In general, the pixel size is adjusted to have about 10 galaxies per pixel. 
If we were having a method to deal with this missing data issue, there would be no particular limitation on the pixel size. However the increasing number of empty pixels would make the mass inversion step always more difficult. Ideally, it would be preferable to have about one galaxy per pixel on average.
%

{\bf Dealing with missing data}\\
Missing data are common practice in weak lensing. They can be due to camera CCD defects, or bright stars that saturate the field of view. More specifically to cluster field reconstruction, the galaxies inside the Einstein radius are usually removed from the study because the weak lensing approximation does not hold there. In addition, depending on the pixel size and the regularity of the galaxy distribution, the amount of empty pixels can increase dramatically. As a result, the measured shear field is generally incomplete and the gaps in the data require proper handling.

A solution that has been proposed by \cite{ejlens:pires09} to deal with missing data consists in filling-in judiciously the masked regions by performing an \emph{inpainting} method simultaneously with a global inversion. Inpainting techniques are an extrapolation of the missing information using some priors on the solution. This new method uses a prior of sparsity in the solution introduced by \cite{ejlens:elad05}. It assumes that there exists a dictionary $\mathcal{D}$ (here the Discrete Cosine Transform) where the complete data are sparse and where the incomplete data are less sparse. The weak lensing inpainting problem consists of recovering a complete convergence map $\kappa$ from the incomplete measured shear field $\gamma_i^{obs}$. The solution is obtained by minimizing
\begin{equation}
\min_{\kappa}  \| \mathcal{D}^T \kappa  \|_0    \textrm{ subject to }   \sum_i \parallel \gamma_i^{obs} - M (P_i * \kappa)   \parallel^2 \le \sigma,
\end{equation}
noting  $|| z ||_0$ the $l_0$ pseudo-norm, i.e. the number of non-zero entries in $z$ and $|| z ||$ the classical $l_2$ norm (i.e. $|| z || =\sum_k (z_k)^2$), where $\sigma$ stands for the  standard deviation of the input shear map, and $M$ is the binary mask (i.e. $M_i = 1$ if we have information at pixel $i$, $M_i = 0$ otherwise).($\sigma=0$ is only used for noiseless data).

If $\mathcal{D}^T \kappa$ is sparse enough, the $l_0$ pseudo-norm can also be replaced by the convex $l_1$ norm (i.e. $ || z ||_1 = \sum_k | z_k | $) \cite{ejlens:donoho_01}. The solution of such an optimization task can be obtained through an iterative thresholding algorithm called MCA \citep{ejlens:elad05} starting from the noisy $\kappa_0$ obtained with the KS93 method
\begin{equation}
\kappa_{i+1} = \Delta_{\mathcal{D},\lambda_n}\left(\kappa_i + M[P_1*(\gamma_1^{obs}-P_1*\kappa_i)+P_2 * (\gamma_2^{obs}-P_2*\kappa_i)]\right),
\label{eqn_mca}
\end{equation}

where the nonlinear operator $\Delta_{\mathcal{D},\lambda}(Z)$ consists in:
\begin{itemize}
\item[-] decomposing the signal $Z$ on the dictionary $\mathcal{D}$ to derive the coefficients $\alpha = \mathcal{D}^T Z$.
\item[-] threshold the coefficients with a hard-thresholding (${\tilde \alpha} = \alpha_i$ if $ | \alpha_i | > \lambda_i$ and $0$ otherwise). The threshold parameter $\lambda_i$ decreases with the iteration $i$.
\item[-] reconstruct $\tilde Z$ from the thresholded coefficients ${\tilde \alpha}$.
\end{itemize}

This method enables to reconstruct a  complete convergence map $\kappa_n$.

\begin{figure}
\centering
\includegraphics[width=\linewidth]{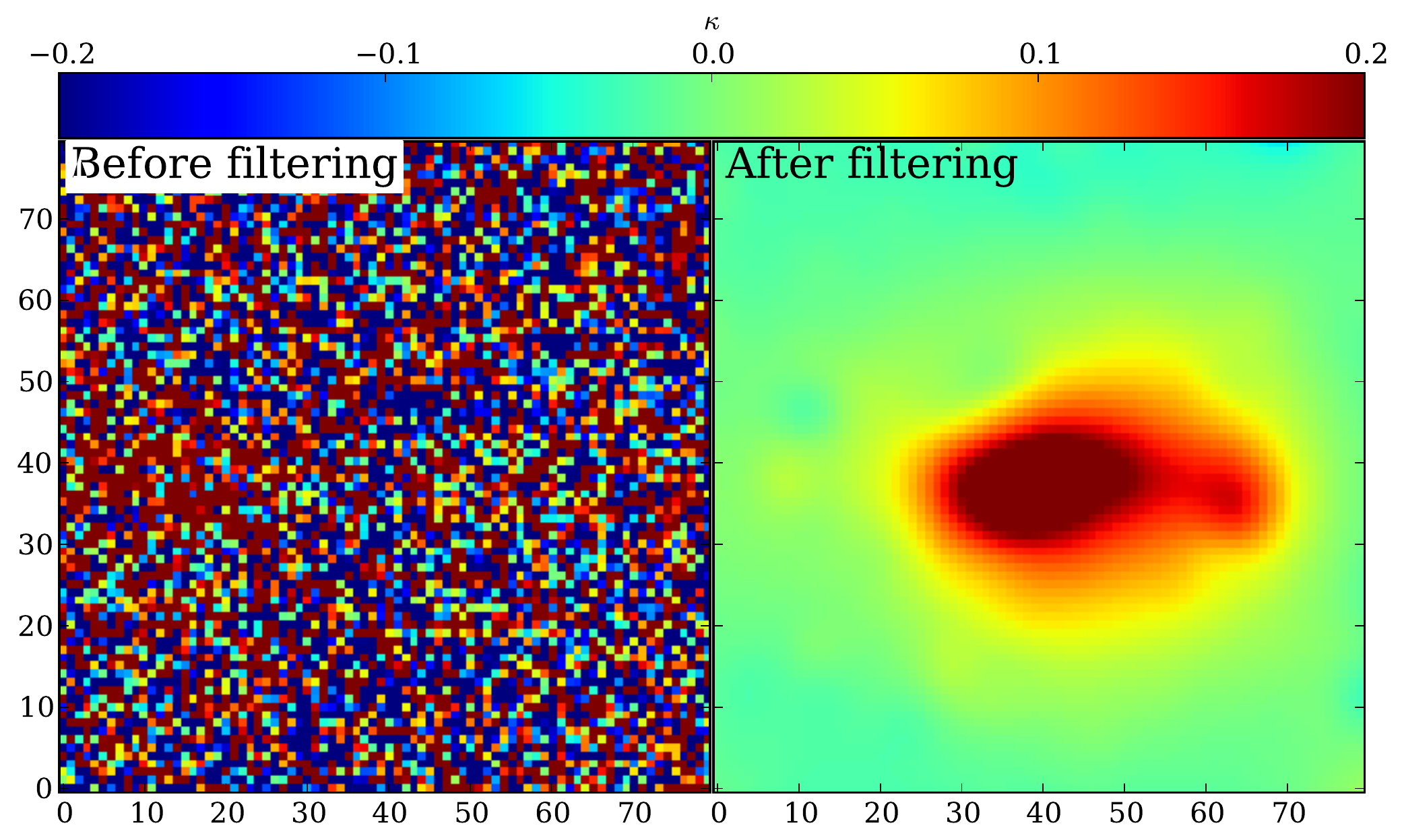}
\caption{Illustration of the filtering of a raw in-painted convergence map with FLens.}
\label{fig:filter}
\end{figure}

However, this  convergence map $\kappa_n$ obtained by inversion of the shear field is very noisy as shown in the left panel of Fig~\ref{fig:filter}. This noise originates from the shear measurement errors and the intrinsic galaxy shape noise, and grows inversely proportional to the number of galaxies per pixel. \\

{\bf Dealing with noise in the Cluster reconstruction}\\
In this study, we use the MRLens (Multi-Resolution for weak Lensing) denoising method to denoise the reconstructed convergence map $\kappa$.
The MRLens filter is based on the Bayesian theory that considers that some prior information can be used to improve the solution \citep{ejlens:starck06}. Bayesian filters search for a solution that maximizes the posterior probability $P(\kappa|\kappa_n)$ defined by the Bayes  theorem :
\begin{eqnarray}
P(\kappa|\kappa_n)=\frac{P(\kappa_n|\kappa)\ P(\kappa)}{P(\kappa_n)},
\label{bayes}
\end{eqnarray}
where :
\begin{itemize}
\item $P(\kappa_n|\kappa)$ is the likelihood of obtaining the data $\kappa_n$ given a particular convergence distribution $\kappa$.
\item $P(\kappa_n)$ is the probability of having the data $\kappa_n$. This term, called evidence, is simply a constant that ensures that the posterior probability is correctly normalized.
\item $P(\kappa)$ is the prior probability of the estimated convergence map $\kappa$. This term codifies our expectations about the convergence distribution before acquisition of the data $\kappa_n$. 
\end{itemize}

Searching for a solution that maximizes posterior probability $P(\kappa|\kappa_n)$ is the same as searching for a solution that minimizes the following quantity
\begin{eqnarray}
\mathcal{Q} &=&  - \log(P(\kappa|\kappa_n)), \\
\mathcal{Q}&=& - \log(P(\kappa_n|\kappa)) - \log(P(\kappa)).
\label{qteinfo1}
\end{eqnarray}
If the noise is uncorrelated and follows a Gaussian distribution, the likelihood term $P(\kappa_n|\kappa)$ can be written
\begin{eqnarray}
P(\kappa_n|\kappa) \propto \exp\ -\frac{\chi^2}{2},
\label{vraisemblance2}
\end{eqnarray}
with the sum of squares of the residuals
\begin{eqnarray}
\chi^2 = \sum_{x,y} \frac{(\kappa_n(x,y)-\kappa(x,y))^2}{\sigma^2_{\kappa_n}}.
\label{vraisemblance3}
\end{eqnarray}

Eq~\ref{qteinfo1} can then be expressed as 
\begin{eqnarray}
\mathcal{Q} =  \frac{1}{2} \chi^2 - \log(P(\kappa)) = \frac{1}{2} \chi^2 - \beta H,
\label{qteinfo2}
\end{eqnarray}
where $\beta$ is a constant that can be seen as a parameter of regularization and 
$H$ represents the prior that is added to the solution.

If we have no expectation about the distribution of the convergence field $\kappa$, the prior probability $P(\kappa)$ is uniform and searching for the maximum of the posterior $P(\kappa|\kappa_n)$ is equivalent to the well-known maximum likelihood search. This maximum likelihood method has been used by \citet{ejlens:bartelmann96} and \citet{ejlens:seljak98} to reconstruct  weak lensing fields, but the solution has to be regularized in some way to prevent overfitting of the data.

Choosing the prior is one of the most critical aspect in Bayesian analysis.
An Entropic prior is frequently used but there are many definitions for Entropy \citep[see][]{ejlens:gull84}. One currently in use is the Maximum Entropy Method (MEM) \citet[i.e.][]{ejlens:bridle98}. A multi-scale maximum entropy prior has also been proposed by \cite{ejlens:marshall02} which uses the intrinsic correlation functions (ICF) with varying width.

The MRLens filtering uses a prior based on the sparse representation of the data that consists in replacing the standard Entropy prior by a wavelet based prior \cite{ejlens:pantin96} . 
The entropy is now defined by
\begin{equation}
 H(I) = \sum_{j=1}^{J-1} \sum_{k,l} h(w_{j,k,l})\;,
\end{equation}
where $J$ is the number of wavelet scales, and we set $\beta = 1$ in Eq.~\ref{qteinfo2}.
In this approach, the information content of an image $I$ is viewed as sum of information at different scales $w_{j}$. The function $h$ defines the amount of information relative to a given wavelet coefficient \cite[see][for details on the choice of this function]{ejlens:starck06}. In \cite{ejlens:pantin96}, it has been suggested to not apply the
regularization on wavelet coefficients which are clearly detected
(i.e. significant wavelet coefficients). The multi-scale entropy then becomes
\begin{eqnarray}
h_n(w_{j,k,l}) =  {\bar M}(j,k,l)  h(w_{j,k,l})  
\end{eqnarray}
where ${\bar M}(j,k,l) = 1 - M(j,k,l)$, and $M$ is the multiresolution support  \cite{ejlens:mur95_2}:
\begin{eqnarray} 
M(j,k,l) = \left\{ \begin{array}{ll} \mbox{ 1 } & 
\mbox{ if
}   w_{j,k,l} \mbox{ is significant} \\ \mbox{ 0 } & \mbox{ if }  
w_{j,k,l}
\mbox{ is not significant} \end{array} \right. 
\end{eqnarray} 
This describes, in a Boolean way, whether the data contains
information at a given scale $j$ and at a given position $(k,l)$.
Commonly, in the case of Gaussian noise, $w_{j,k,l}$ is said to be significant if
$| w_{j,k,l} | > k\sigma_j$, where $\sigma_j$ is the noise standard deviation at scale
$j$, and $k$ is a constant, generally taken between 3 and 5.

The False Discovery Rate method (FDR) offers an effective way to select this constant $k$ \citep{ejlens:benjamini95,ejlens:miller01,ejlens:hopkins02}. 
The FDR defined  as the ratio
\begin{eqnarray} 
FDR = \frac{V}{D}
\end{eqnarray}
where $V$ is the number of pixels erroneously identified as pixels with signal, and $D$ is the number of pixels identified as pixels with signal, both truly and erroneously.

This method requires to fix a rate $\alpha$ between 0 and 1. And it ensures that {\it on average}, the FDR will not be bigger than $\alpha$
\begin{eqnarray} 
E(FDR) \leq \frac{T}{V}.\alpha \leq \alpha
\end{eqnarray}
The unknown factor $\frac{T}{V}$ is the proportion of truly noisy
pixels.  A complete description of the FDR method can be found in
\cite{ejlens:miller01}.  Here we apply the FDR method at each wavelet scale, which gives us a detection threshold $T_j$
per scale. We then consider a wavelet coefficient
$w_{j,k,l}$ as significant if its absolute value is larger than $T_j$. This procedure is totally different from a $k\sigma$ thresholding, that only controls the ratio between the number of pixels erroneously identified over the total number of pixels in the map.

The proposed filter called MRLens (Multi-Resolution for weak Lensing\footnote{The MRLens denoising software is available at the following address: "http://irfu.cea.fr/Ast/mrlens software.php".}) outperforms other techniques (Gaussian, Wiener, MEM, MEM-ICF) in the reconstruction of dark matter. For this reason, it has also been used to reconstruct the dark matter mass map from the Hubble Space Telescope in the COSMOS field \cite{ejlens:massey07}. \\

{\bf Dealing with reduced shear}

In practice, the observed galaxy ellipticities, however, are induced not by the shear $\gamma$ but by the reduced shear
\begin{equation}
g = \frac{\gamma}{1-\kappa}.
\label{eq:redshear}
\end{equation}
The distinction between the true and the reduced shear is negligible in the weak shear regime ($\kappa \approx 0$). However
 in galaxy cluster fields, as we focus on in the work, the weak shear regime is not perfectly satisfied, and the discrepancy in the reconstructions can be as high as 10 \% if the reduced shear is not properly taken into account. 

In order to recover the true shear from the measured reduced shear, we consider an iterative algorithm. At the first iteration,  we assume that the true shear is equal to the reduced shear. Then a  convergence map is derived, and used along with Eq~\ref{eq:redshear} to compute a more accurate true shear for the next iteration. We found this procedure to effectively correct for the bias in the reconstruction, but found no improvement after three iterations.





\subsection{The multi-scale grid model}

\subsubsection{RBF Model description}

Radial Basis Functions (RBFs) are commonly used to solve interpolation problems \citep[see e.g.][]{ejlens:gentile12}. Let us consider an unknown function $f:  \mathbb{R}^n \to \mathbb{R}$ probed  at a set of locations $\xi \in \mathbb{R}^n$, and approximated by a function $s :  \mathbb{R}^n \to \mathbb{R}$, a linear combination of translates of  a set of RBFs $\phi_i$

\begin{equation}
\label{eq:rbf}
s(\bold{x}) = \sum \lambda_i\ \phi_i(|| \cdot - \bold{x}||)\,.
\end{equation} 

\noindent with unknown real coefficients $\lambda_i$. Those coefficients are obtained by solving the linear system $f(\xi) = s(\xi)$. A unique solution exists if there are as many RBFs as data points and the RBF profiles are positive definite \citep{ejlens:buhmann03}. However in our case, since data points are noisy and we want to avoid overfitting, we arbitrarily restrict the number of RBFs to a few, thus practically compressing the data to a smaller basis set. 
 
In \citet{ejlens:jullo09}, we found that RBFs  distributed on a hexagonal grid, and described by a Truncated Isothermal Mass Distribution (TIMD) \citep[see e.g.][]{ejlens:kassiola93,ejlens:kneib96,ejlens:eliasdottir09} were giving good results.
In our model, we approximate the true convergence field $\kappa$ with 
\begin{equation}
\label{eq:conv}
\kappa(\theta) = \frac{1}{\Sigma_{crit}} \sum_i \sigma_i^2\ f(\ || \theta_i -\ \theta\ ||,\ s_i,\ t_i)
\end{equation}

where the RBFs on grid nodes $\theta_i$ are described by  
\begin{equation}
\textstyle
\label{eq:timd}
f(R,s, t) = \frac{1}{2G} \frac{r_{cut}}{t- s} \left( \frac{1}{\sqrt{s^2 + R^2}} - \frac{1}{\sqrt{t^2 + R^2}} \right).
\end{equation}

In the TIMD model, the scaling factor  $\sigma_i^2$ is the velocity dispersion at the centre of the gravitational potential, and radii $s$ and $t$ mark 2 changes in the slope respectively from $\kappa \propto R^0$ to $\kappa \propto R^{-1}$ and $\kappa \propto R^{-3}$ respectively. 

In a similar manner, we approximate the true shear field with
\begin{eqnarray}
\gamma_1(\theta) = \sum \sigma_i^2\ \Gamma_1(\ || \theta_i -\ \theta\ ||,\ s_i,\ t_i)\\
\gamma_2(\theta) = \sum \sigma_i^2\ \Gamma_2(\ || \theta_i -\ \theta\ ||,\ s_i,\ t_i)
\end{eqnarray}

\noindent  where analytical expressions also exist for $\Gamma_1$ and $\Gamma_2$ \citep[see Eq\ A8 in ][]{ejlens:eliasdottir09}.

Let us now consider a set of $M$ ellipticity measurements ordered in a vector $\bold{e} = [\bold{e_1},\ \bold{e_2}]^\dagger$, and a model made of $N$ RBFs distributed in the field with unknown weights $\sigma_i^2$ ordered in a vector $\bold{v} = [\sigma_1^2, \ldots, \sigma_N^2]$.  In the weak lensing approximation, we can write the linear relation

\begin{equation}
\label{eq:shearmat}
\bold{e} = M_{\gamma v} \bold{v} + \bold{n}\;,
\end{equation}

\noindent where $\bold{n}$ is the galaxy shape noise as in Eq~\ref{eq_gamma}, and the transform matrix $M_{\gamma v} = \left[ \Delta_1, \Delta_2 \right]^\dagger$ is a block-2 matrix.  Its individual elements are the contribution of each unweighted RBF scaled by a ratio of angular diameter distances 

\begin{eqnarray}
\label{eq:dshear1}
\Delta_{1}^{(j,i)} &= &\frac{D_{LSi}}{D_{OSi}}\ \Gamma_{1}^i(|| \theta_i - \theta_j ||,\ s_i,\ t_i) , \\
\Delta_{2}^{(j,i)} &= &\frac{D_{LSi}}{D_{OSi}}\ \Gamma_{2}^i(|| \theta_i - \theta_j ||,\ s_i,\ t_i) . 
\end{eqnarray}

\noindent where subscript $j \inÊ[1, M]$ and $i\in [1, N]$ denote the rows and the columns of $M_{\gamma \nu}$ respectively.

\subsubsection{Comparison of TIMD and Gaussian filters}

\begin{figure}
\centering
\includegraphics[width=\linewidth]{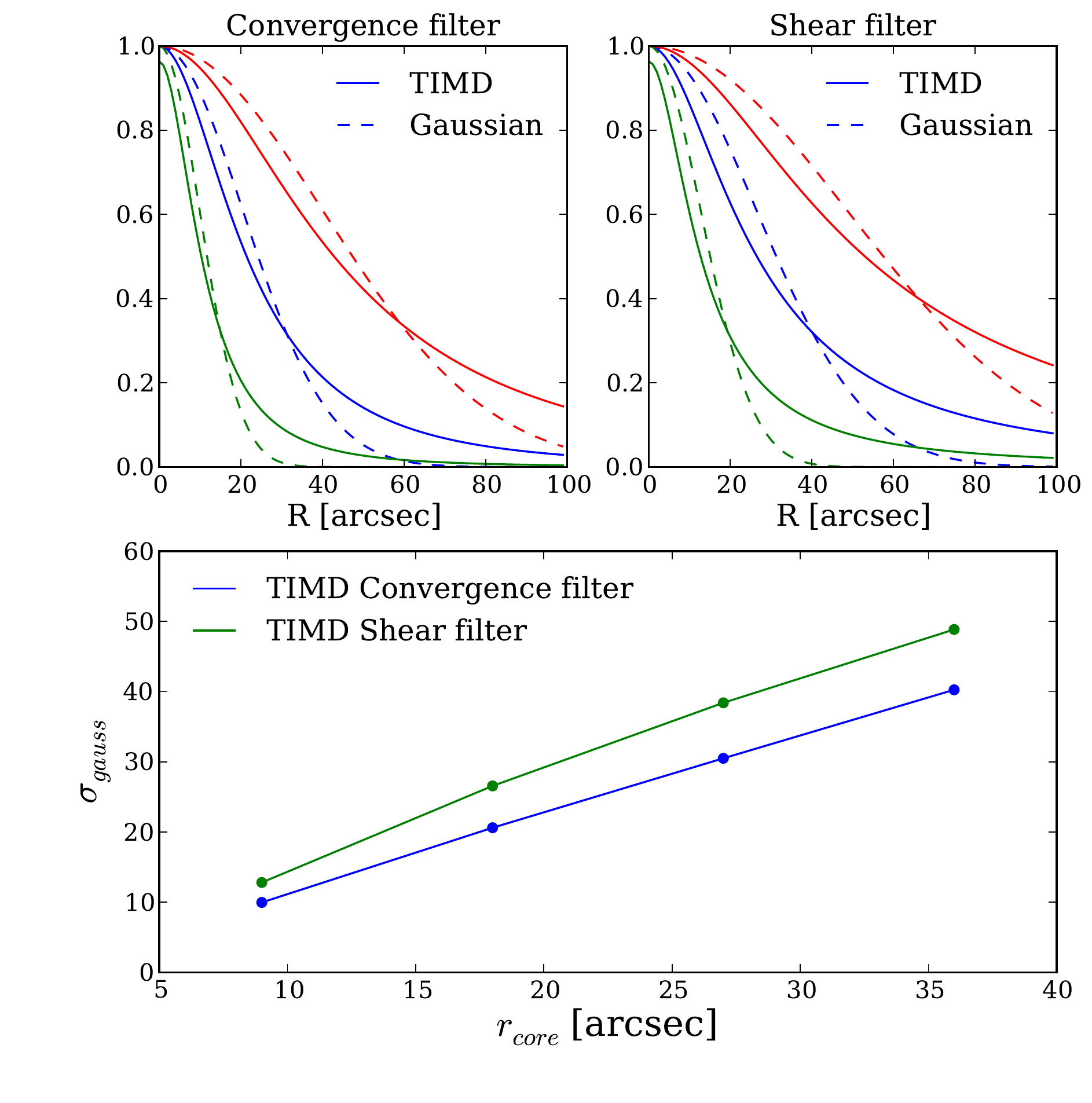}
\caption{Comparison between the TIMD profiles in convergence and shear spaces. In dashed-line, we also show the best-fit Gaussian profiles. The bottom panel shows that the TIMD profile in shear space is systematically broader than its equivalent in convergence space, in comparison to a self-similar Gaussian filter. }
\label{fig:piemdgauss}
\end{figure}


By construction, we use the same parameters for the RBFs ($\sigma^2_i,\ s_i,\ t_i$) in the convergence and shear spaces. However, the corresponding functions $f$, $\Gamma_1$ and $\Gamma_2$ have different profiles in these two spaces. In Fig~\ref{fig:piemdgauss}, we actually show that the TIMD filter is sharper in convergence space than in shear space. In practice, this makes the TIMD filter very efficient at picking shear information far away for a given RBF, and concentrate it to produce high resolution convergence maps. For example from Fig~\ref{fig:piemdgauss} we see that if we use a TIMD filter of core radius $s= 20"$ (equivalent to a Gaussian filter of width $\sigma \simeq 30"$ in shear space), the reconstructed convergence field is smoothed similarly as with a Gaussian filter of width $\sigma \simeq 22"$. In contrast with the standard KS93 method, the size of the Gaussian filter is the same in shear and convergence space.

\subsubsection{Estimation of the RBFs weights}

{\bf Linear SVD inversion method}

Assuming the galaxy shape noise $\bold{n}$ is Gaussian distributed, we can write the sum of the squares of the residuals
\begin{equation}
\label{eq:chi2}
\chi^2 = (\bold{e} - 2 M_{\gamma v} \bold{v})^\dagger N_{e e}^{-1} (\bold{e} - 2 M_{\gamma v} \bold{v}),
\end{equation}

\noindent where $N_{e e} \equiv\ < e e^\dagger >$ is the covariance matrix of the measured ellipticities. In this work, we assume this matrix is diagonal and its elements are $N_{e e}^{(i,j)} =(\sigma_m^2 + \sigma_{int}^2 )\ \delta_{ij}$ where $\delta_{ij}$ is the Kronecker symbol,  $\sigma_m$ is the measurement uncertainty and $\sigma_{int}$ is the scatter in the distribution of the intrinsic shapes of the galaxies. Note also that we have a factor of $2$ in this equation because in Lenstool the ellipticity $e = \frac{a^2 - b^2}{a^2 + b^2}$ is computed as a function of the square of the major and minor axes   \citep{ejlens:bartelmann01}.  
With Gaussian distributed errors, linear inversion theory tells us that an unbiased estimator of the RBF weights is 
 \begin{equation}
 \tilde{\bold{v}} = \left[ M_{\gamma v}^\dagger N_{e e}^{-1} M_{\gamma v} \right]^{-1} M_{\gamma v}^\dagger N_{e e}^{-1}  \bold{e}
 \end{equation}
 
\noindent and their covariance is  
\begin{equation}
\label{eq:nvv}
N_{v v}  =  \left[ M_{\gamma v}^\dagger N_{e e}^{-1} M_{\gamma v} \right]^{-1} 
\end{equation}

The convergence field is obtained by the matrix product
\begin{equation}
\tilde{\kappa} = M_{\kappa v}\, \tilde{v}
\end{equation}

\noindent and the corresponding covariance matrix $N_{\tilde{\kappa} \tilde{\kappa}}$  by  
\begin{equation}
\label{eq:svdconv}
N_{\tilde{\kappa}Ê\tilde{ \kappa}} = M_{\kappa v} N_{v v} M_{\kappa v}^\dagger\;.
\end{equation}

\noindent where the transform matrix $M_{\kappa v}$ is built from Eq~\ref{eq:conv} and \ref{eq:timd}. In the following, we reconstruct the convergence field in grids of regularly spaced pixels.

There are several ways of speeding the calculations in the expressions above. In particular, it happens that in our case, the transform matrix $ M_{\gamma v}$ is sufficiently sparse so that we can perform a singular value decomposition (SVD). Details of the SVD decomposition can be found in \citep{ejlens:vanderplas11,ejlens:diego05}.\\

{\bf Bayesian MCMC optimisation}

\begin{figure}
\centering
\includegraphics[width=\linewidth]{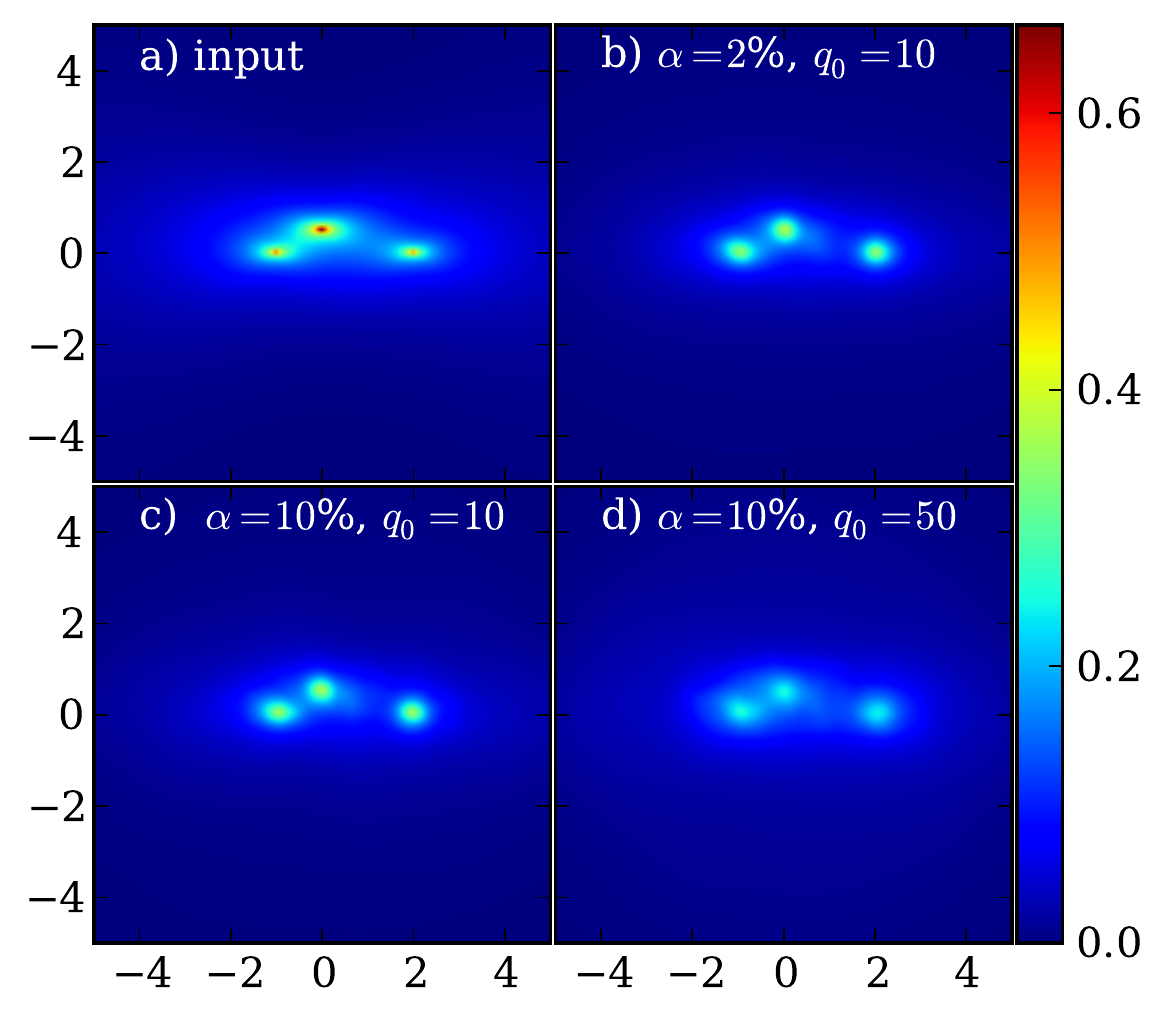}
\caption{Impact of different user defined nuisance parameters on the Lenstool reconstruction of a simulated convergence map. Parameter $q_0$ has the strongest impact on the reconstruction result. These reconstructions are without shape noise, and with a multi-scale grid of 575 RBFs. }
\label{fig:alphacomp}
\end{figure}

In this section, we describe the Bayesian Monte Carlo Markov Chain algorithm used to reconstruct the mass map in \citet{ejlens:jauzac12}. This algorithm called MassInf is also part of the Bayesys package \citep{ejlens:jullo07}, but it is the first time we use it in Lenstool\footnote{Lenstool public package is available at the following address http://projects.lam.fr/projects/lenstool}. 
It aims at inverting linear systems of equations in a Bayesian manner, i.e. with input priors. 

Based on our definition of the $\chi^2$ in Eq~\ref{eq:chi2}, we define the likelihood of having a set of weights $\bold{v}$ given the measured  ellipticities $\bold{e}$ as
\begin{equation}
\label{eq:lhood}
P(\bold{v}\, |\,Ê\bold{e}) = \frac{1}{Z_{L}} \exp{- \frac{ \chi^2}{2}}.
\end{equation}
The normalization factor is given by $Z_{L} = \sqrt{(2 \pi)^{2M} \det N_{e e}}$.

As a prior, we want the individual weights $\sigma_i^2$ to be positive, so that the final mass map is positive everywhere. This conducted us to assume they are described by a Poisson   probability distribution function (pdf)
\begin{equation}
\mathrm{Pr}(\sigma_i^2) = \exp(-\sigma_i^2 / q) / q\,,
\end{equation}

\noindent where the normalization factor $q$ is a nuisance parameter with a pdf given by the following expression
\begin{equation}
\pi(q) = q_0^2 q e^{-q/q_0}\,.
\end{equation}

This expression has been chosen to be tractable analytically whilst keeping $q$ away from 0 and $\infty$. The parameter $q_0$ is fixed and seeded by the user. In our case, we found that $q_{0} = 10$ was giving good performances in terms of computation time, and reconstruction fidelity against the simulated data. In Fig~\ref{fig:alphacomp}, we show that its exact value has little impact on the final reconstruction.

In contrast to the standard Bayesys algorithm implemented in Lenstool, Massinf does not explore all the correlations between the parameters, but searches for the most relevant parameters (keeping the others fixed meanwhile), and explores their PDF individually, reproducing thus somehow the Gibbs sampling approach. It also makes use of an additional nuisance parameter called $n$, which is the number of RBFs the sampler estimates necessary to reproduce the data. We obtained good results with this number described by a geometric pdf
\begin{equation}
\mathrm{Pr}(n) = (1 - c) c^{n-1}  \quad \mathrm{where} \quad c = \frac{\alpha}{\alpha + 1}\,,
\end{equation}

\noindent and parameter $\alpha = 2\%$ of the total number of RBFs. Again we show in Fig~\ref{fig:alphacomp} that this parameter has little impact on the reconstruction.

\section{Simulated filament study}
\label{sec:simus}

\begin{figure}
\centering
\includegraphics[width=\linewidth]{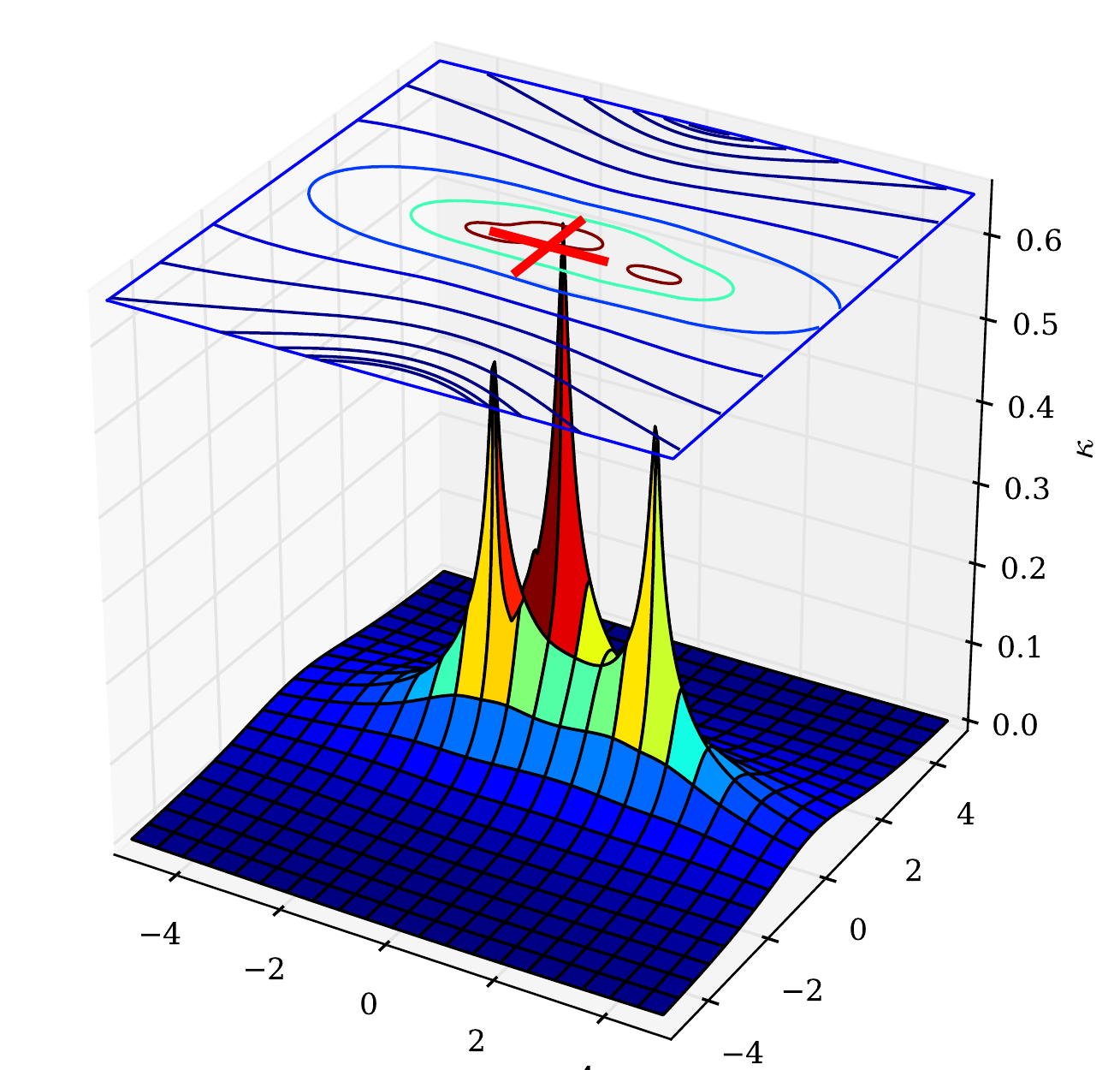}
\caption{Simulated filamentary structure with 3 elliptical NFW clumps. The cross indicates the center of the field. Contour levels are in log scale between $10^{-4} < \kappa < 0.2$. }
\label{fig:input3d}
\end{figure}

We applied our reconstruction algorithms to a simulated mass map made of 3 NFW halos at redshift $z=0.5$. The field of view is $10 \times 10$ square arcminutes, and the 3 halos are located at (0, 0.5'), (-1', 0) and (2', 0) in equatorial coordinates. They form a 3' long filamentary structure aligned along the right ascension axis. To emphasize the extended aspect of the structure we made the halos elliptical with an ellipticity $e = \frac{a^2 - b^2}{a^2 + b^2} = 0.4$. For each halo, the scale radius is $r_s = 300$\ kpc (50''), and their concentration are $c = 3$ and $c=3.5$ for the halo central halo. This translates into masses $M_{200} = 1.4\times 10^{14} M_{\odot}$\ and $M_{200} = 2.3\times 10^{14} M_{\odot}$ in a $\Lambda$CDM cosmology $(\Omega_m = 0.3, \Omega_\Lambda =0.7, H_0 = 70\ {\rm km.s^{-1}.Mpc^{-1}}, w_0 = -1)$.\\

From this mass model, we generated a convergence map by setting the sources at redshift $z= 1.2$, which is reasonable for data coming from the Hubble Space Telescope, alike the COSMOS data. This convergence map is shown in Fig~\ref{fig:input3d}. We also produced  reduced shear catalogs with sources taken randomly across the field of view, and to which we added a random intrinsic ellipticity drawn from a Gaussian pdf of width $\sigma_{\rm int} = 0.27$. Again, this is a reasonable value for data coming from HST \citep{ejlens:leauthaud07}.\\

\subsection{Standard galaxy density  catalog}
\label{sec:model}

\begin{figure}
\centering
\includegraphics[width=\linewidth]{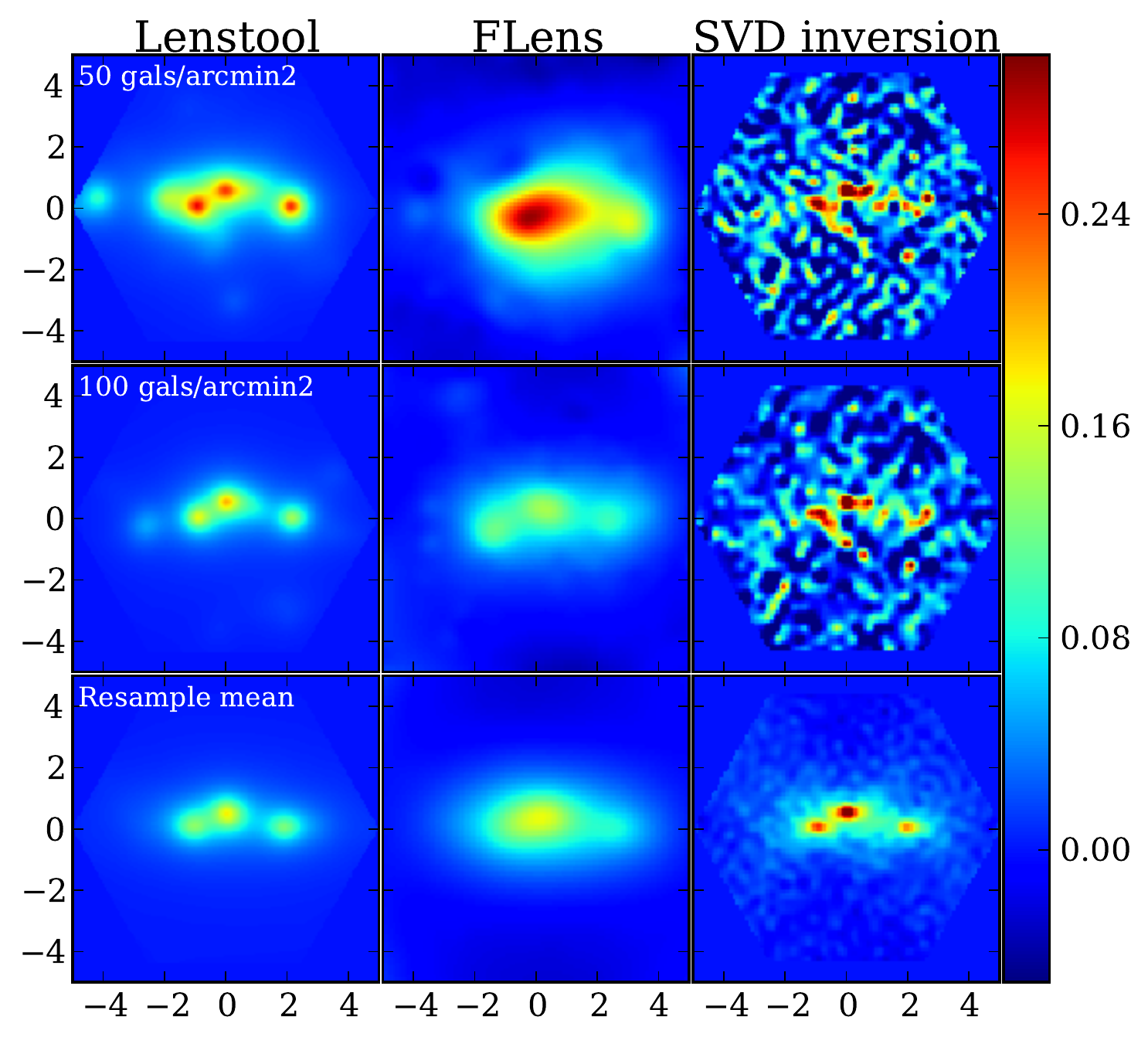}
\caption{Convergence maps reconstructed with the three methods. Top panel reconstructions are made with 50 gals/arcmin$^2$, middle and bottom panels with 100 gals/arcmin$^2$. Bottom panel is obtain after resampling 100 times the shape noise of the input catalog. Globally, Lenstool and FLens reconstructions have a lower noise level than SVD reconstruction. Lenstool reconstructions have high resolution, but also contain spurious peaks, whereas FLens reconstructions have lower resolution, but no spurious peaks. Resampling is efficient at removing the spurious peaks in all 3 cases and increases the signal to noise of the SVD reconstructed peaks.}
\label{fig:comparison}
\end{figure}

First, we compare the reconstruction obtained with a catalog containing 5,000 sources, i.e. with a density of 50 galaxies per square arcminute. Results are shown in the top panel of Fig.~\ref{fig:comparison}. At first, we note that Lenstool and FLens produce less noisy reconstructions than the SVD inversion method. The Lenstool reconstruction has high resolution, but also contains spurious peaks, whereas the FLens reconstruction has lower resolution, but no spurious peaks. 

In this simulation, the FLens map is 64x64 pixels, and the pixel size is 0.156'. To filter out the reconstructed noise, Flens uses a wavelet decomposition procedure that only keeps scales with $J>3$, i.e. structures larger than 8 pixels in size. As described in \ref{sec:flens}, this wavelet scales thresholding is controlled by the FDR method. If a scale is noise dominated, the detection threshold will be very high and the scale will be removed, thus degrading the resolution of the reconstructed map. This global estimation of the detection threshold per scale is more robust to the noise but less sensitive to small structures.
A more local approach would increase the resolution and the detection of small structures, but would also increase the number of false detections.

For the Lenstool and SVD inversion methods, we adjust the resolution of the grid-based reconstruction to the power spectrum of the input signal. Peaks can still be resolved by  cutting high frequencies  at $k > 10$ arcmin$^{-1}$ ($k = \frac{2\pi}{R}$). This translates into RBFs with core radius $s = 0.3'$. We choose an hexagonal grid of RBFs in order to limit high frequency noise at the junction between nearby RBFs. We can cover the whole FOV with a grid of 817 RBFs. The Lenstool reconstruction is less noisy than the SVD reconstruction essentially because of the priors implemented in Lenstool. 

\subsection{High galaxy density catalog}

In order to increase the resolution of the FLens reconstruction, we produce a catalog with 10,000 sources, i.e. with a 100 galaxies per square arcminute. Results are shown in the middle panel of Fig.~\ref{fig:comparison}.  By doubling the size of the catalog, we could decrease by 4 the pixel size (0.04'), and detect the halo on the right in the FLens reconstructed map. The Lenstool reconstruction still contains spurious peaks.\\

\subsection{Shape noise resampling}


In the two previous analysis, we observed some overfitting of the galaxy shape noise,  especially with Lenstool and the SVD inversion, leading to spurious peaks. \\

In order to mitigate this issue, we resample 100 times the intrinsic galaxy shape noise in the input catalog of 10,000 sources. We run Lenstool, FLens and the SVD reconstructions on each of 100 catalogs, and average the reconstructed convergence maps. The outcome of this procedure is presented in the bottom panel of Fig.~\ref{fig:comparison}. \\

We note that the spurious peaks have disappeared from the averaged maps, but also that the power in the peaks is globally less than in the original map.\\

\subsection{Reconstructed density profile}

\begin{figure}
\centering
\includegraphics[width=\linewidth]{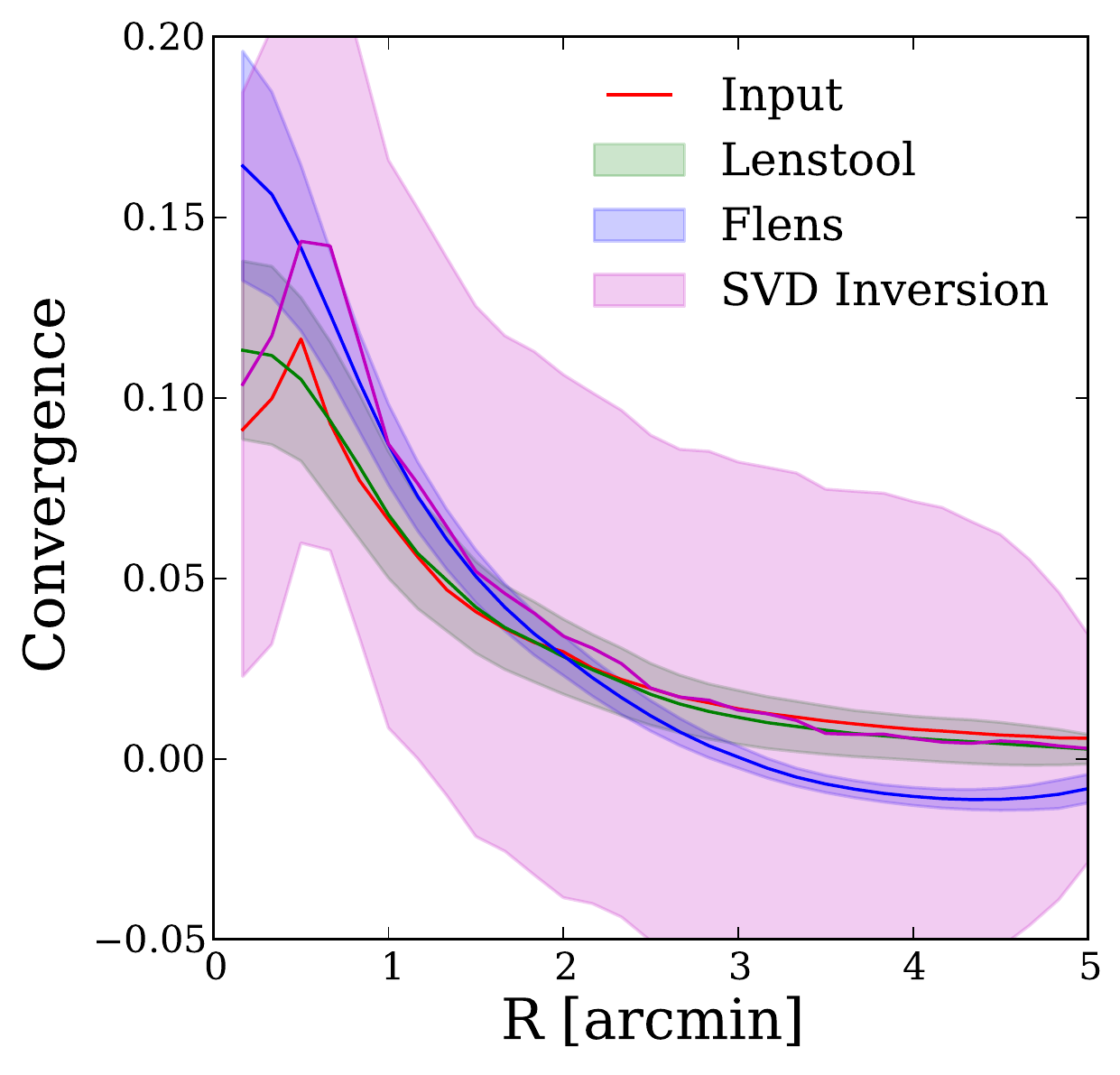}
\caption{Comparison of the convergence profile recovered with FLens,  Lenstool and the SVD inversion, assuming 100 galaxies per sq. arcmin., and resampling of the noise. Errors are given at 68.2\% C.L.}
\label{fig:simu_profile}
\end{figure}

It is a very common procedure in galaxy cluster studies to average the reconstructed mass maps azimuthally to produce a radial density profile. We perform this measurement for our three methods and compute the errors by taking the standard deviation of the 100 reconstructed maps. 

In Fig.~\ref{fig:simu_profile}, we show the comparison of the azimuthally averaged density profiles. The striking point of this figure is the amount of noise in the SVD reconstruction. The second point is the fact that the FLens density profile becomes negative at radius $R > 180$ arcsec and over-estimates the density at small radius. This is due to the fact that wavelets are compensated filters with null mean. In contrast, Lenstool reconstruction is unbiased, and contains the input profile in its 1$\sigma$ confidence contours. The correct normalization at large radius is due to the fact that Lenstool takes into account the redshifts of the lens and the individual sources in the fit.\\

\subsection{Errors on the reconstructed maps}

\begin{figure}
\centering
\includegraphics[width=\linewidth]{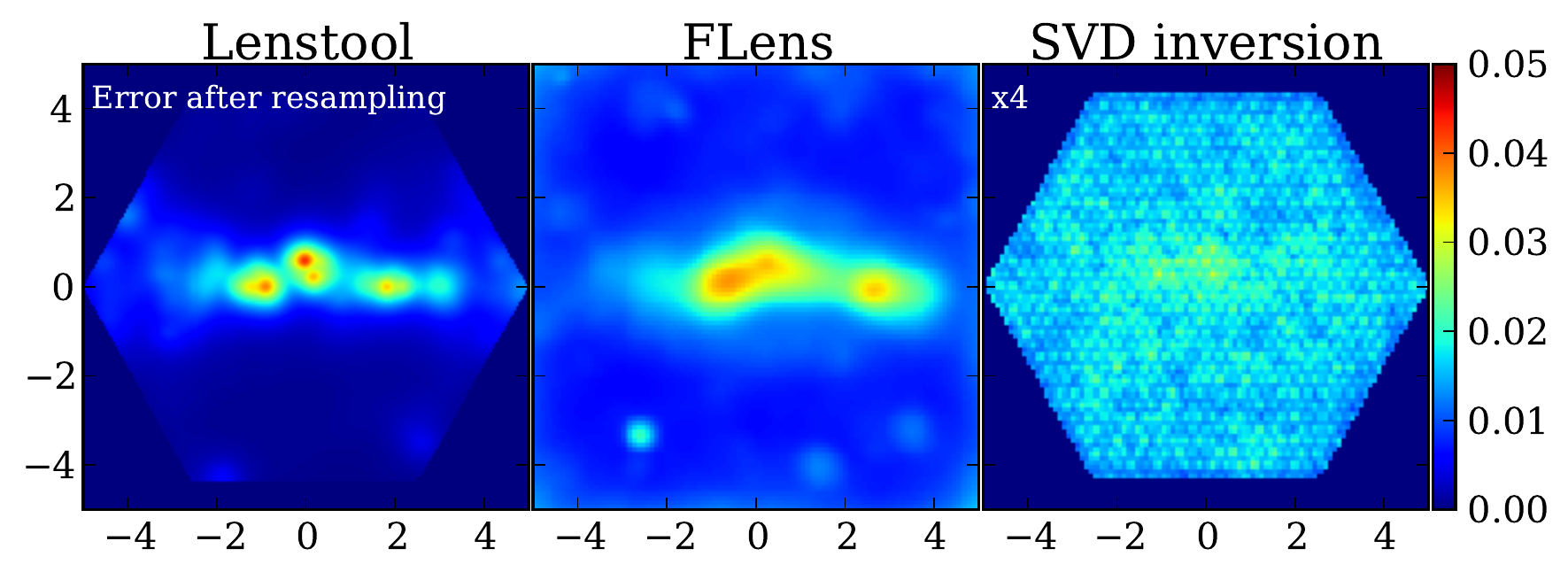}
\caption{Errors on the reconstructed convergence maps with the three methods. In theory SVD errors are independent of the underlying shear signal, but we still notice that locally they depend on the galaxy density. SVD errors have been divided by 4 to fit the colormap range.}
\label{fig:stdcomp}
\end{figure}


We compute the errors of the reconstructed maps following 2 approaches. The Lenstool and the SVD inversion methods output an estimate of the error in each pixel, either by means the analysis of the MCMC samples, or the covariance matrix computed in Eq~\ref{eq:svdconv} respectively. Nonetheless to get rid of overfitting, we resample the galaxy shape noise in the input catalogs, and compute the variance of the pixels reconstructed both with Lenstool, Flens and the SVD inversion. Fig~\ref{fig:stdcomp} show that with the three methods, the errors scale with the input density field.

It is worth noticing that the SVD error map also scales with the input signal, although the covariance matrix  $N_{\kappa \kappa}$  does not directly depend on the ellipticity measurements $\bold{e}$. We have done some tests, and found that with a uniform distribution of galaxies, this effect vanishes. \emph{Therefore, it seems this effect is due to lensing amplification}, which decreases the amount of galaxies in this region, and as a result increases the variance in the reconstruction.

Finally, we have found that using RBFs with larger core radius increases the correlations between the RBF weights in $N_{vv}$, and decreases the resolution, as well as the overall signal to noise. In contrast, using RBFs with smaller core radius produces higher resolution but  noisier reconstructions. We found that matching size of the RBFs to the grid resolution yields the best compromise.\\

\subsection{Errors on the reconstructed density profiles}

\begin{figure}
\centering
\includegraphics[width=\linewidth]{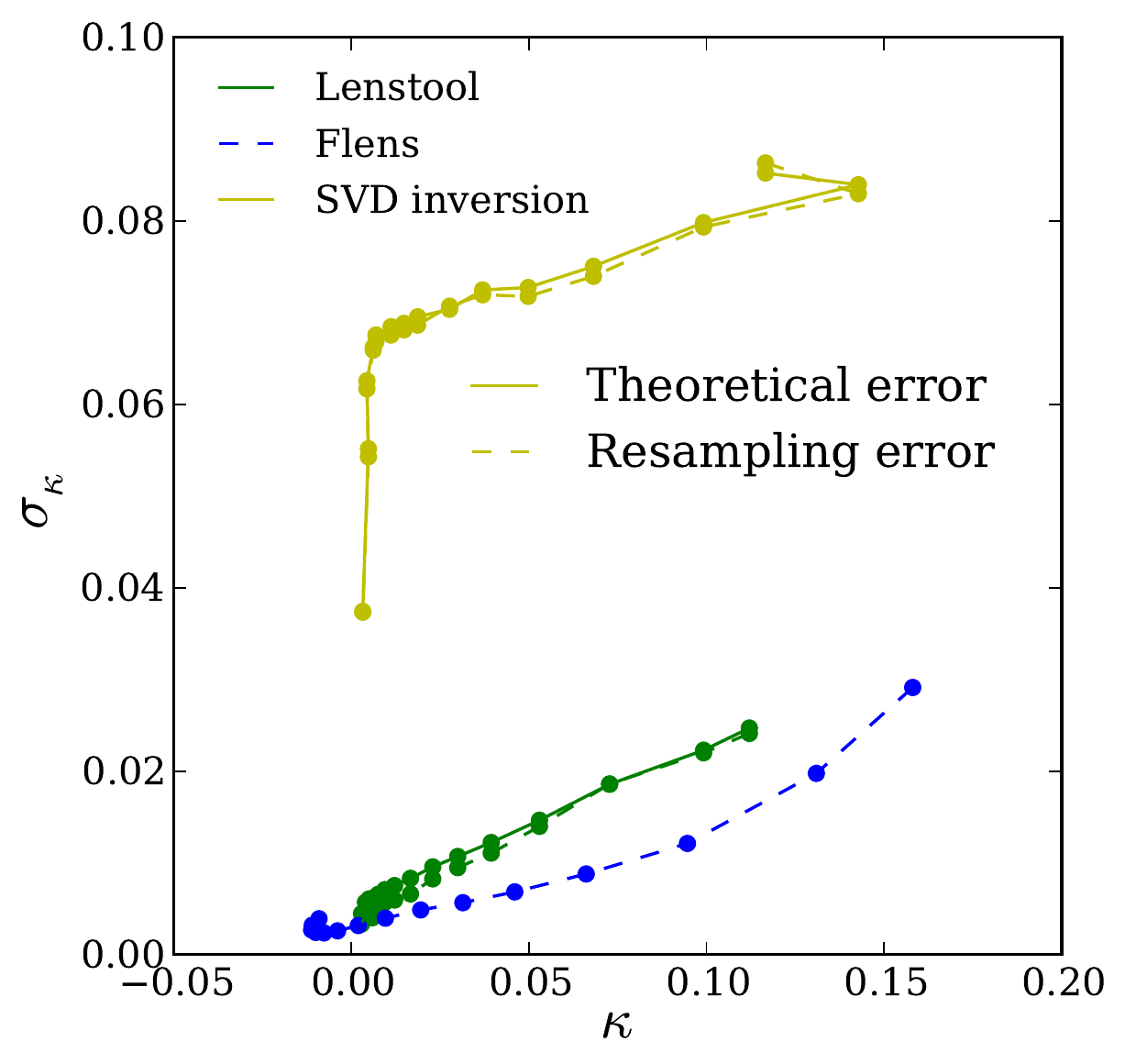}
\caption{Scaling of reconstructed noise as a function of reconstructed signal for different reconstruction methods. SVD inversion and Lenstool methods both provide a way to directly estimate errors on the reconstruction. This is what we call \emph{Theroretical errors}. These errors are in good agreement with errors estimated with noise resampling.}
\label{fig:sigkok}
\end{figure}

We then focus on the estimated errors on the azimuthally averaged density profiles. In Fig~\ref{fig:sigkok}, we find that the errors scale with the reconstructed density, in agreement with what we observed in the errors on the reconstructed maps. With this figure, we clearly see that the SVD inversion produces errors about 4 times larger than what can be achieved with Lenstool or FLens methods.

Besides, it is reassuring to see that the errors estimated from the Lenstool MCMC samples or the covariance matrix $N_{\tilde{\kappa}Ê\tilde{\kappa}}$ agree with errors estimated after resampling.

Regarding the bias between the reconstructed and the true convergence profiles, we note from Figure~\ref{fig:simu_profile} that Lenstool bias is almost constant at less than 5\% from the input values, whereas FLens  and SVD biases increase with $\kappa$ and reach about 30\% at $\kappa = 0.07$.

\subsection{Signal to noise estimates}

\begin{figure}
\centering
\begin{tabular}{c}
\includegraphics[width=0.85\linewidth]{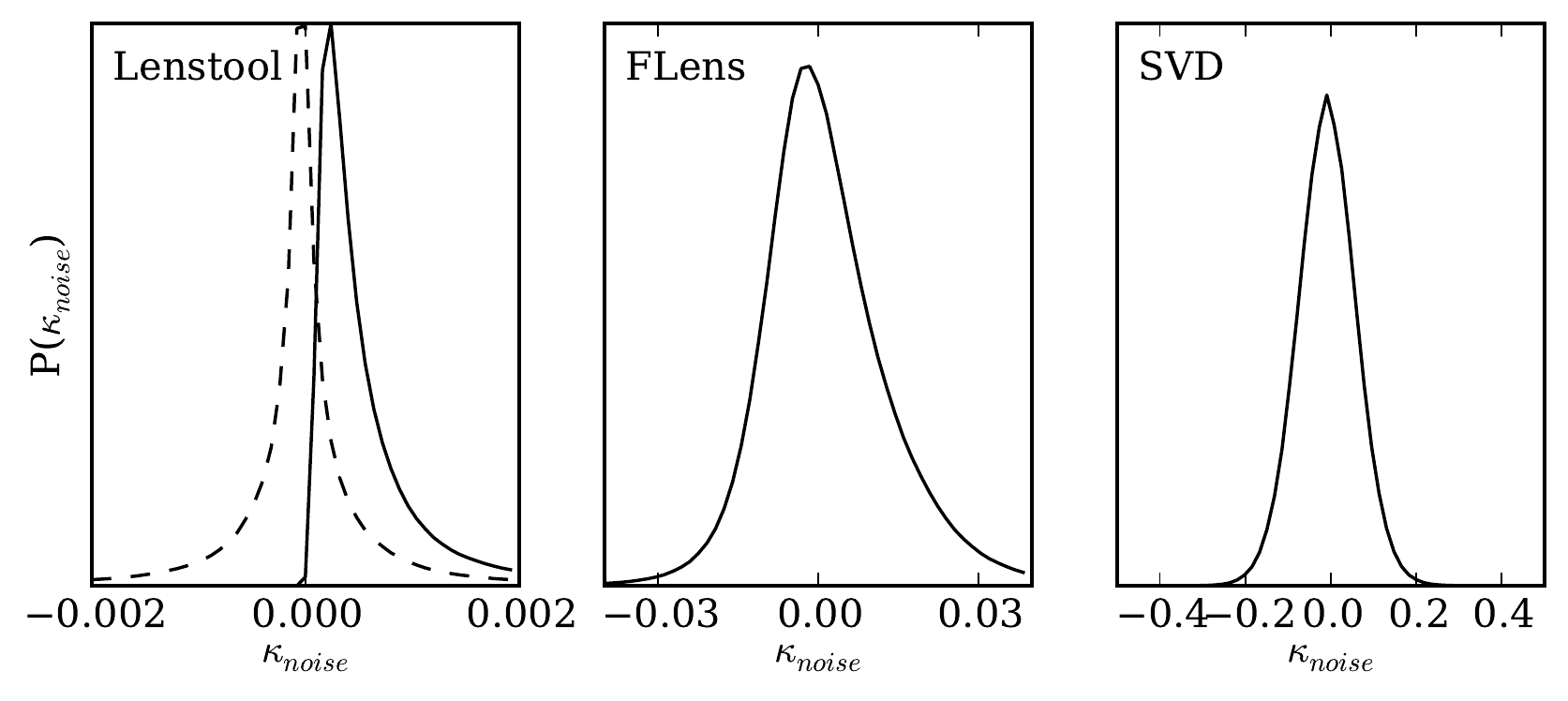} \\
\includegraphics[width=\linewidth]{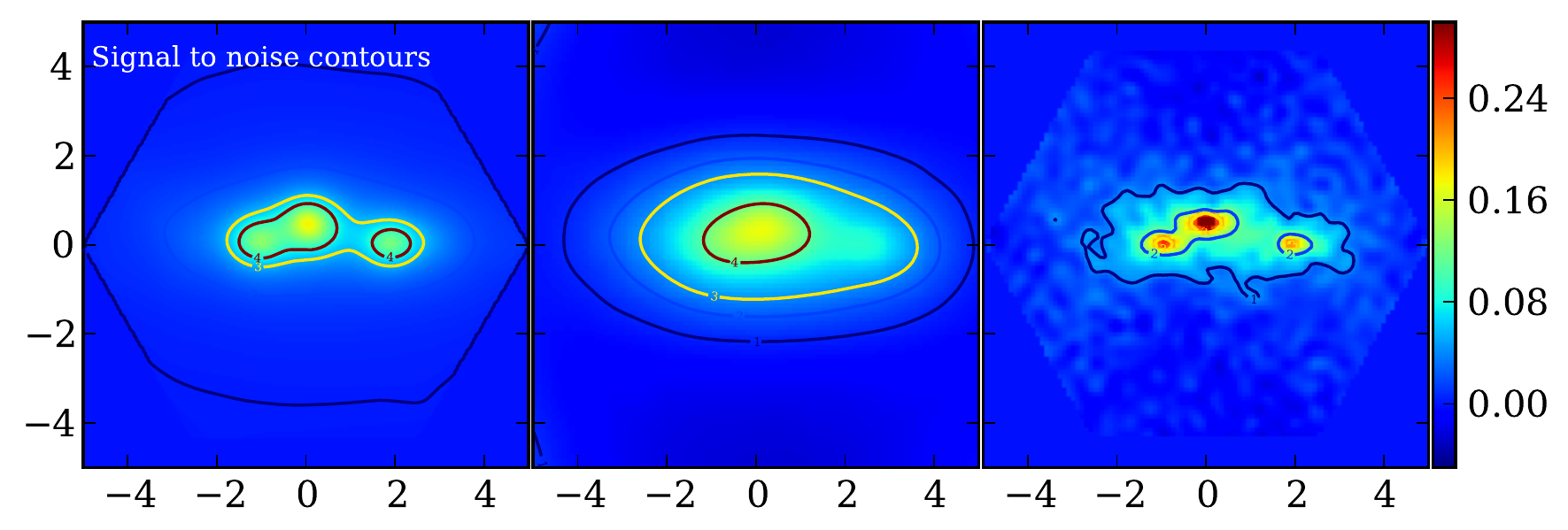}
\end{tabular}
\caption{{\bf Top panel} From left to right, probability distribution functions of the convergence reconstructed from 1000 noise maps, as obtained with Lenstool, FLens and the SVD inversion respectively. The dashed curve corresponds to Lenstool without the prior on positive convergence.  {\bf Bottom panel} Reconstructed convergence maps with 100 galaxies per sq. arc-minutes and noise resampling. Contours indicate the levels of confidence at 68.2\%, 95.5\%, 99.7\% and 99.9\%. }
\label{fig:snmap}
\end{figure}

It is a common procedure to compute the signal to noise by dividing the estimated signal by the variance of the noise. However in the top panel of Fig.\ref{fig:snmap}, we show that in our case, the pdf of the reconstructed noise is not necessarily Gaussian distributed. This is particularly evident for the Lenstool method. 

From each pdf, we therefore compute the threshold X, for which we have the probability of finding a value $x$, $P(x \le X)$ equals to 68.2\%, 95.5\%, 99.7\% and 99.9\%. We found that with 1000 realizations of noise, we had enough statistics to estimate up to only 4$\sigma$ level.

In the bottom panel of Fig.\ref{fig:snmap}, we observe that the SVD inversion is more noisy than the Lenstool or the FLens methods. The 1$\sigma$ region of the confidence is larger with Lenstool and smaller with the SVD inversion. Globally, the regions of equal confidence are similar in size with Lenstool and FLens, especially at larger signal to noise.

\section{Application to MACSJ0717+3745}
\label{sec:macs0717}

In this section, we apply our three methods to the real case of the galaxy cluster MACS J0717+3745, in which a filament was recently detected with Lenstool multi-scale grid reconstruction \citep{ejlens:jauzac12}. 

\subsection{Modeling description}

The analysis in \citet{ejlens:jauzac12} was based on a mosaic of 18 multi-passband images obtained with the Advanced Camera for Surveys aboard the Hubble Space Telescope, covering an area of $\sim10\times 20$ square arcminute.
The weak-lensing pipeline developed for the COSMOS survey, modified for the analysis of galaxy clusters, was used to produce a weak-lensing catalogue of roughly 52 galaxies per square arc-minutes. A uBV color diagram was used to distinguish the background sources from the foreground and cluster-member galaxies. Their redshift distribution was derived from photometric and spectroscopic redshifts obtained from Subaru and CFHT/WIRcam imaging in the same field \citep{ejlens:ma08}. Because they are in the strong lensing regime area, all the galaxies inside an elliptical region of 5 x 3 arc-minutes in size and 45$^\circ$-rotated centered on the cluster core were also removed from the catalog. The detail of the catalog construction is thoroughly described in \citet{ejlens:jauzac12}.

In order to compute error bars on the reconstructions, we resampled the weak lensing catalog with a bootstrap strategy, i.e. each galaxy in the catalog can be removed or duplicated, in order to increase its weight in the reconstruction. We produced 50 of such bootstrapped catalogs.

For the Lenstool and the SVD inversion methods, we built a grid of RBFs. In contrast to the model described above, in which all the RBFs had the same size, in Jauzac et al. we used a multi-scale grid with smaller RBF in regions where the cluster luminosity was brighter. First, we built a smoothed cluster luminosity map from the catalog of magnitudes in K-band of cluster member galaxies. Then, we computed a multi-scale grid of RBFs, making sure that the luminosity in each triangle was lower than a predefined threshold. As a result, we obtained a grid made of 468 RBFs, the smallest ones having a core radius $s = 26$ arcsec. 

\subsection{Reconstructed maps}

\begin{figure*}
\centering
\includegraphics{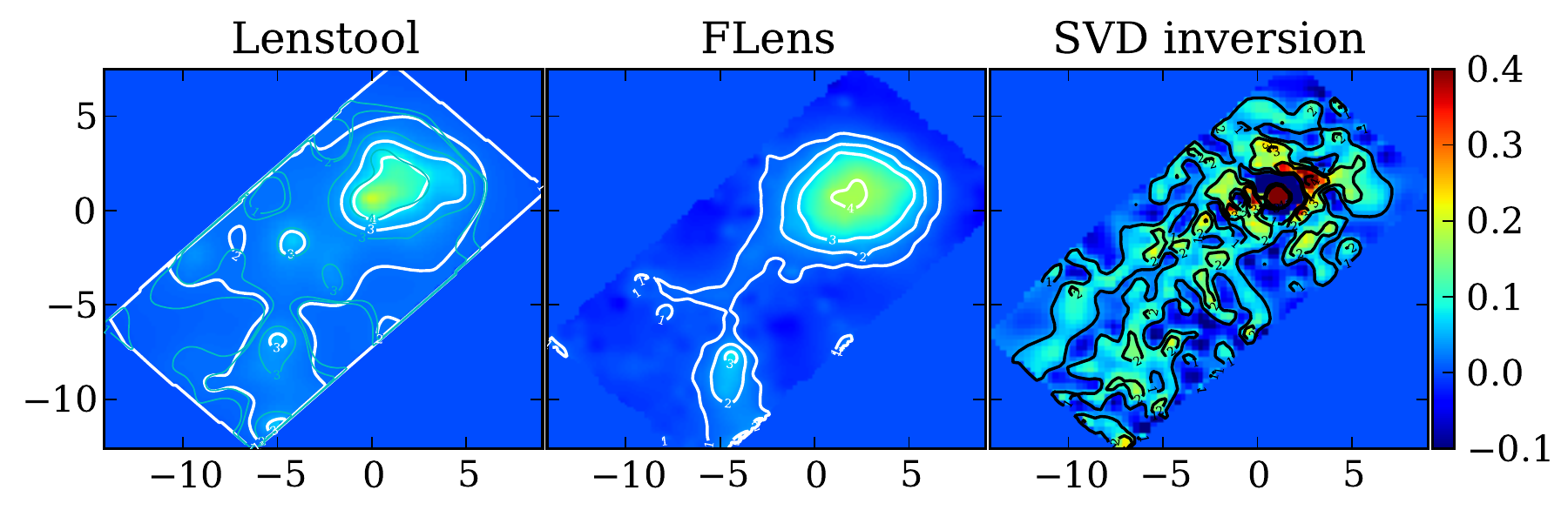}
\caption{Reconstructed convergence maps of MACS J0717 with the three methods. Signal to noise contours are based on 1000 noise maps computed by randomizing the ellipticities of the galaxies in the input catalog. They assess the level of confidence of the detected structures at 68.2\%, 95.5\%, 99.7\% and 99.9\%. Cyan contours in the Lenstool panel correspond to a reconstruction without the prior of positive convergence. North is up, East is left. Coordinates are in arc-minutes relative to the cluster center $\alpha = 109.39102$ and $\delta = 37.746639$}.
\label{fig:comp0717}
\end{figure*}

Fig~\ref{fig:comp0717} shows the reconstructed convergence maps of MACS J0717 obtained with the three methods. Globally, they all agree on the location of the cluster core, and the presence of an extension to the South-East. In the cluster core where data are missing, both the Lenstool and FLens reconstructions are smooth, whereas the SVD reconstruction is more clumpy. We attribute this difference to the priors assumed in both Lenstool and FLens.

We also observe some disagreement on the exact shape of the filament. 
Lenstool reconstruction suggests MACS J0717 lies into an extended over-dense region. In contrast, FLens reconstruction shows that the cluster is compact and connected to a long filament.  In both the FLens and the Lenstool reconstructions, the filament is detected at 95\% C.L. The SVD reconstruction presents 2 filaments next to each other.

\subsection{Reconstructed density profile}

\begin{figure}
\centering
\includegraphics[width=\linewidth]{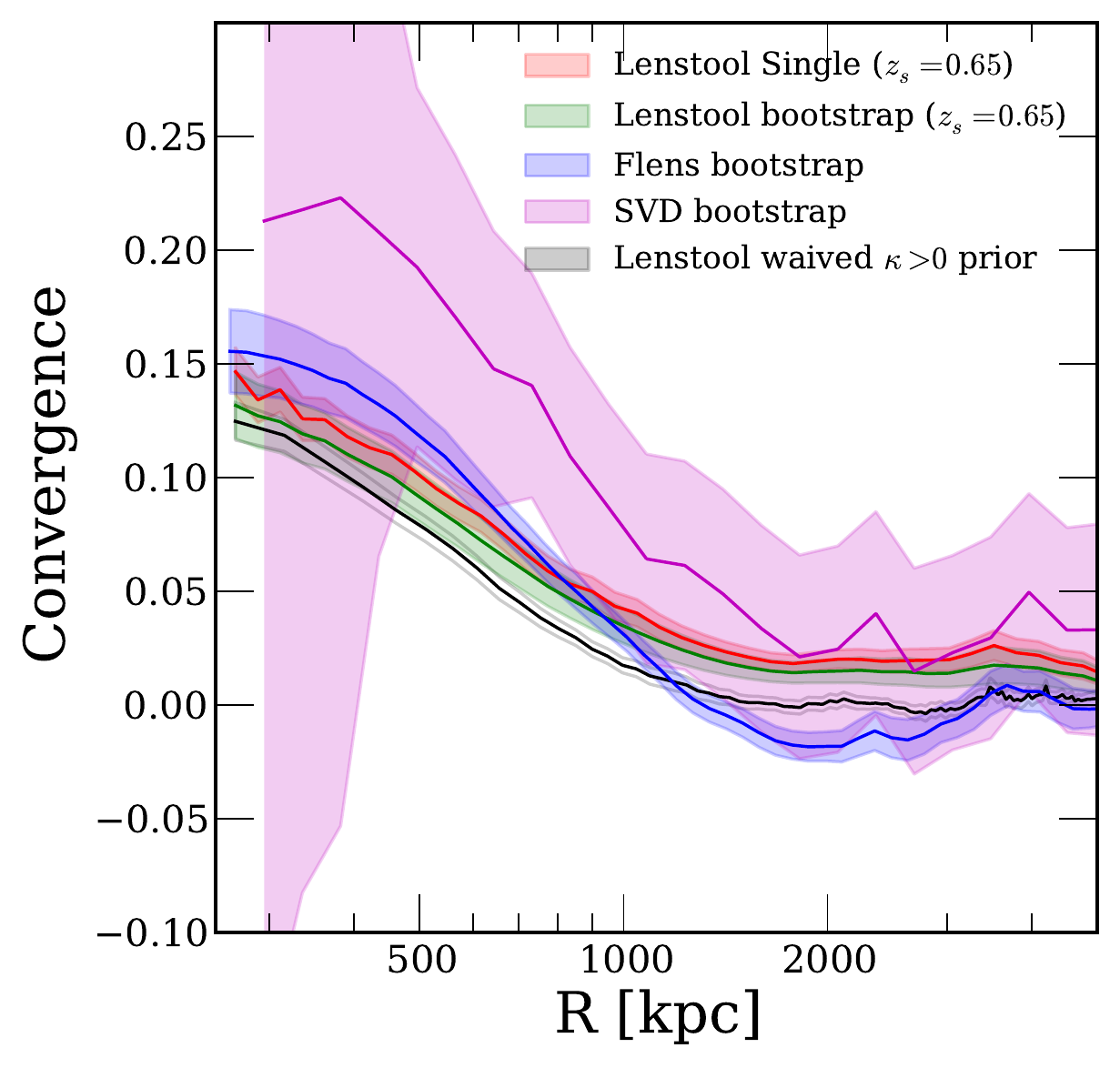}
\caption{Reconstructed convergence profiles obtained with the three methods. The FLens profile is in good agreement with the other profiles in the core, but deviates at large radius, were it becomes negative. The SVD reconstruction is not able to reproduce the central high density region. Without the positive prior on $\kappa$, we obtain a better agreement between Lenstool and FLens at large radius.}
\label{fig:sdens}
\end{figure}

Fig.\ref{fig:sdens} shows the corresponding radial convergence profiles obtained with the three methods. We took the coordinates $\alpha = 109.39102$ and $\delta = 37.746639$ as the central point of the azimuthal average. Based on the photometric redshift analysis performed in \citet{ejlens:jauzac12}, we assumed in the Lenstool reconstruction that the redshift of the weak lensing sources to be $z_s = 0.65$. 

As already observed in the simulations, the noise level estimated from bootstrap is about 4 times larger in the SVD reconstruction, especially close to the cluster center. The Lenstool method agrees with FLens at small radii, and with the SVD inversion at large radius. The FLens method predicts steeper radial profile between 500 kpc and 1000 kpc, and a bump at 3 Mpc, corresponding to the over-density in the filament. This feature is much less evident in the other reconstructions. 

Note as well that the convergence profiles derived from single catalog and bootstrap catalogs reconstructions with Lenstool agree together. Lenstool error estimates from the MCMC sampling are therefore reliable.

\section{Conclusion}

Systematic errors in lensing map reconstruction, especially due to the reconstruction methods, is a concerning issue. With the current and forthcoming datasets, they start to dominate the error budget over the statistical errors. 

In this work, we have studied three methods of reconstruction of 10 arc-minutes scale structures, i.e. the environment of galaxy clusters. We limited our study to a toy-model structure in order to focus on the effect of priors. In a forthcoming paper, we will increase the level of complexity by using N-body simulations. The FLens method starts from a pixelated map of shear, with about one galaxy per pixel on average, and filter the noisy reconstructed convergence map by only keeping wavelet scales that contain non Gaussian signal. 

The Lenstool and the SVD inversion methods share the same underlying multi-scale grid model. The field is paved with a set of RBF, whose number density and size scale with the smoothed surface brightness of the cluster member galaxies. Lenstool uses a Bayesian MCMC sampler to estimate the weight of each RBF in the reconstruction, where the SVD inversion makes use of the linear formalism of the weak-lensing approximation to estimate the weights. The RBF shape is defined from the Truncated Isothermal Mass Distribution (TIMD), which can either give the shear for the inversion or the convergence for the reconstruction. 

So far with Lenstool, we have forced the density field and therefore the convergence to be positive everywhere. This assumption is valid here, because we consider the case of massive structures. Nonetheless in order to be exhaustive in this study, we also turned this prior off in Lenstool and redid all the computations. We found very similar results both for the simulated case and for MACSJ0717.
\\

From the simulations, we found the following~:
\begin{itemize}
\item All three methods can detect clusters and surrounding filaments in the convergence range $0.01 < \kappa < 1$, although with different levels of significance.

\item Doubling the galaxies number density from 50 to 100 per square arcminute allows to reduce the pixel size and increase the resolution of the FLens reconstruction. The resolution of Lenstool and the SVD inversion methods is more driven by the density of RBFs than by the galaxy density. However, the signal to noise per pixel increases with galaxy number density.
 
\item The error on the reconstructed convergence scales with the underlying signal, and depends on the method used for the reconstruction. The residual is offset from zero by a small amount, that decreases when we increase the grid resolution. 

\item Thanks to the inpainting technique implemented in FLens, we could recover the shape of the cluster even in reasonably high density regions ($\kappa \sim 0.16$). 

\item We compared these results to the forward fitting method presented in \citet{ejlens:jauzac12} and implemented in \textsc{Lenstool}. The forward fitting method recovers the true density map with deviations less than 5\% at $\kappa > 0.5$, and less than 20\% at $0.5 > \kappa > 0.01$. In contrast to the other method, the redshift of the cluster and sources are used as a constraint to break the mass-sheet degeneracy. As a result no significant offset is found in the residual. 

\item We found FLens to be more robust against shape noise than Lenstool or standard inversion methods. Resampling techniques increase the signal to noise of regions with low signal, but decrease signal to noise of regions with high signal.
\end{itemize}

We applied the new method to the galaxy cluster MACSJ0717, and confirmed the presence of the filament at 3$\sigma$ C.L. We also repeated the Lenstool analysis previously done in \citet{ejlens:jauzac12}, but this time with a bootstrap of the input source catalog. The consistent results obtained with these two techniques give us more confidence in the detection of the structures around MACSJ0717. Without the prior of positive convergence applied, we obtained a very similar map and consistent signal to noise contours in Figure~\ref{fig:comp0717} , and a density profile in better agreement with FLens at large radius in Figure~\ref{fig:sdens}. 

To conclude, it is very encouraging to see that priors can significantly enhance the signal to noise in weak lensing reconstructions. FLens priors are strictly limited to the properties of the galaxy shape noise. In contrast, Lenstool priors enforce the mass-follows-light assumption to build the multi-scale grid. Ideally, the science goals condition the type of priors to choose. A weak lensing peak counting analysis to characterize dark energy might prefer limited priors in order to better compare to theory, whereas the exploration of the cosmic web might heavily rely on external priors coming from other observables, such as galaxy density, X-ray or SZ maps.


\vspace{1cm}
The authors would like to thank J.-L. Starck for useful discussions. Computations have been performed at the M\'esocentre d'Aix-Marseille Universit\'e. This work was supported by the European Research Council (ERC) grant SparseAstro (ERC-228261). JPK acknowledges support from the ERC advanced grant LIDA and from CNRS.

\bibliographystyle{aa}
\bibliography{ejlens}

\label{lastpage}

\end{document}